\documentclass[12pt]{article}
\usepackage[utf8]{inputenc}
\usepackage{geometry}
\usepackage{setspace}
\usepackage{titlesec}
\usepackage{graphicx}
\usepackage{hyperref}
\usepackage{float}

\title{Trust in Transparency: How Explainable AI Shapes User Perceptions}
\author{Allen Daniel Sunny \\
University of Maryland, College Park \\
\texttt{allens@umd.edu}}
\date{}

\begin{document}

\maketitle

\begin{abstract}
This study explores the integration of contextual explanations into AI-powered loan decision systems to enhance trust and usability. While traditional AI systems rely heavily on algorithmic transparency and technical accuracy, they often fail to account for broader social and economic contexts. Through a qualitative study, I investigated user interactions with AI explanations and identified key gaps, including the inability of current systems to provide context. My findings underscore the limitations of purely technical transparency and the critical need for contextual explanations that bridge the gap between algorithmic outputs and real-world decision-making. By aligning explanations with user needs and broader societal factors, the system aims to foster trust, improve decision-making, and advance the design of human-centered AI systems.
\end{abstract}


\section{Introduction}
We live in an age of technological acceleration. The telephone took 78 years to reach 50 million users, while radio took 38 years.\cite{interactive_schools_50_million} In stark contrast, ChatGPT achieved the same milestone in just two months. \cite{tooltester_chatgpt_statistics}. This rapid adoption underscores a seismic shift in how society integrates new technologies. Historically, the ability to absorb economic and social changes caused by technological breakthroughs has taken time, allowing for trust to build naturally through iterative use and understanding \cite{un_desa_technological_acceleration} However, the accelerated pace at which AI systems are being deployed has disrupted this historical pattern. Today, we see AI systems being incorporated into critical decision-making roles such as finance\cite{murawski_mortgage_ai}, college admissions\cite{pangburn_schools_ai}, criminal justice\cite{springer_ai_governance} and the banking industry.\cite{deloitte_ai_banking}

To safety adopt technology, you need to trust it.\cite{choung_trust_ai} A critical component of this trust-building process is explainability. \cite{Bussone2015}\cite{Roszel2021} For previous technological breakthroughs, this trust-explainability relationship was fostered through transparency and accountability. Engineers understood the systems they built and could explain them to stakeholders.\cite{mcknight_trust_technology} Today, with the rise of opaque AI models, this foundational understanding is often missing. Even the engineers behind these systems frequently lack insight into how decisions are made, leaving end-users and stakeholders blind to critical processes.\cite{Hoffman2021}

Responding to this issue, the field of Explainable AI (XAI) has progressed at a rapid clip. \cite{nguyen_xai_review} It has given us algorithmic approaches to generate explanations of how an AI system behaves and makes decisions.\cite{Chuang2023} The field has, however, been criticized for having a myopic focus.\cite{green_algorithmic_realism}  \cite{selbst2019fairness} As most XAI techniques are designed by XAI researchers for other XAI researchers, it excludes those who are actually affected by the AI system. \cite{Aechtner2022} Research has shown that the explanations preferred by XAI researchers do not translate well to other individuals who are actually affected by these systems.\cite{SmithRenner2020} \cite{Bach2024} As with most poorly deployed systems, it tends to affect the most marginalized within society.\cite{selbst2019fairness} Explanations as a construct are usually more effective when looked at through a socio-technical lens. Explanation is first and foremost a shared meaning-making process that occurs between an explainer and an explainee. This process is dynamic to the goals and changing beliefs of both parties.\cite{Ehsan2020}

In response to these criticisms, there have been several studies, including \cite{Rong2023},  \cite{Yang2017}, \cite{Kouki2019}that have attempted to extend user centric AI development in order to develop paradigms that can help define certain benchmarks in creating AI systems that are able to build trust. This has helped create a new field, Human-AI Interaction (HAI). This new subfield has tackled the problem of building trust by bridging work between AI, HCI and work in critical theory such as \cite{agre1997computation} that have given insights into this issue. Attempts have been made in order to conclusively design frameworks that address the explainability-trust framework. \cite{Jacovi2021} attempted to use critical theory to analyze and critique the social and ethical implications of explanations generated by AI systems. This approach emphasizes the role of context and power dynamics in determining what constitutes a "reasonable" explanation. On the other hand, \cite{miller2019explanation} provided a foundational understanding of explanations by leveraging insights from cognitive and social sciences, focusing on how humans generate and interpret explanations in contrastive and causal terms. Building on these perspectives, \cite{Nauta2023} defined the COP-12 metrics, which offer a structured framework for evaluating the quality of AI explanations. These have broad themes of Content, Presentation and User Focus. 
These metrics emphasize key dimensions such as clarity, relevance, fidelity, and utility, ensuring that explanations are not only accurate but also meaningful and actionable for end-users.

In spite of these efforts, most of the research in the HAI space does not touch users that have no context of the system they are going to be affected by. Most current studies such as \cite{Ehsan2020} use experts in their respective fields. Surveys of papers such as \cite{Nauta2023}\cite{Rong2023} are still biased towards expert opinions. In this paper, we look at the current explanation paradigms in the HAI field and attempt to see if they still hold weight when evaluated by end-users who have very limited understanding of AI systems as well as minimal contextual knowledge. Or more broadly, how useful is the current state-of-the-art in explanations within the HAI field in this new context? In summary, our contributions are:
\begin{itemize}
    \item RQ1: How do different types of AI explanations (e.g., example-based, rule-based) shape users’ initial impressions and perceived trust during their interaction with the AI system?
    \item RQ2: What elements of these explanations are most influential in building or undermining users’ trust during their interactions with a AI system?
    \item Reaffirming whether the current metrics used to evaluate HAI are effective in diverse real-world contexts and what perspectives are missing
\end{itemize}

\section{Related Work}
We will start with a review of the XAI field and how a new sub field called Human AI Interaction (HAI) has arisen. We'll also review the benefits of looking at AI development through a sociotechnical scale. 

\subsection{Explainable AI (XAI)}
Explainable AI (XAI) aims to uncover how AI models behave and to communicate these behaviors to users in a way that fosters understanding. The nature of XAI work varies depending on the specific objectives, whether it's diagnosing model behavior or providing actionable insights to end-users. Researchers in this field often develop diagnostic explanations, designed to help practitioners and developers identify and address model issues. In contrast, actionable explanations are more suited to non-technical users, enabling them to make informed decisions without requiring a deep understanding of the model's inner workings.\cite{Hoffman2021}

To address the needs of broader audiences, XAI research has introduced post-hoc explanations, such as counterfactual explanations \cite{Le2023}, which aim to elucidate model behavior without relying on diagnostic tools. These explanations attempt to bridge the gap for users who lack the technical capability to interpret complex model outputs. However, recent findings have shown that explainability is audience-dependent rather than model-deterministic; what resonates with one user may fail to convey meaningful insights to another. This highlights the critical need for tailoring explanations to specific user groups.\cite{Cai2019}

Despite these advancements, significant challenges persist in XAI research. Many common XAI techniques have been tested only in controlled laboratory settings with user research \cite{velez2017interpretable}, limiting their generalizability. In practice, these techniques often struggle to effectively convey explainability to diverse user populations. Moreover, the relationship between explainability and user trust has shown mixed results, with XAI techniques frequently falling short in fostering trust among users.\cite{choung_trust_ai} \cite{Duarte2023} Achieving both explainability and trustworthiness in real-world systems remains a persistent challenge.

Another issue lies in the sociotechnical aspect of XAI. Much of the current research focuses on the deployment of AI systems in closed business environments, analyzing their behavior within these constrained settings. There is limited exploration of how XAI systems perform when deployed at scale in large, open environments, where diverse user needs and societal implications come into play.

\subsection{Human-AI Interaction (HAI)}
Current HAI techniques focus on improving user interaction and fostering trust through mechanisms such as explainability, interpretability, and personalized design.\cite{olah2018building} Methods like Explainable AI (XAI) aim to make AI systems more transparent by providing insights into how decisions are made, while techniques such as interactive visualizations and natural language explanations enhance user engagement and understanding. Personalization is increasingly emphasized, with AI systems adapting to individual user needs, contexts, and expertise levels. \cite{fogarty2017personalized} However, these approaches face significant critiques. Many explanations remain too complex, failing to account for diverse user capabilities and cognitive loads.\cite{lipton2016mythos} The "black-box" problem persists, with opaque algorithms making it difficult for even developers to fully understand AI decision-making processes. Furthermore, current HAI techniques often lack empirical grounding in psychological or sociological research, resulting in explanations that may not align with how users naturally process information.\cite{miller2019explanation} Ethical concerns, such as potential bias in explanations and the manipulation of user perceptions, also highlight the need for critical evaluation and more robust, inclusive design frameworks. As AI systems increasingly take on high-stakes roles, addressing these critiques is essential for building trustworthy, human-centered solutions.\cite{sweeney2013discrimination}
Although HAI is growing in a more inclusive direction, the vast amount of it's research is still focused around the computer science or domain specific spaces. Wide ranging surveys such as \cite{Nauta2023} \cite{Li2023}\cite{Rong2023} \cite{Nguyen2024} all showcase this concentration.
In particular \cite{Nauta2023} from which the COP-12 metrics were derived shows that of the nearly 400 papers under review, 22\% of them only contain user studies. Out of that , only 23\% of the remaining are focused on non-domain experts. This skews reported metrics  which fails to take into account perspectives that are completely outside of the space. 

\subsubsection{COP-12}
The COP-12 framework, introduced by Nauta et al\cite{Nauta2023}., is a structured set of metrics designed to evaluate the quality of AI explanations. It identifies 12 key dimensions grouped into three main categories: Content, Presentation, and User-Focused Metrics. Content metrics (e.g., correctness, completeness, consistency) assess the factual accuracy and comprehensiveness of explanations. Presentation metrics (e.g., compactness, composition) evaluate how explanations are structured and delivered to users. Finally, user-focused metrics (e.g., context, coherence, controllability) emphasize the alignment of explanations with user needs and their ability to interact with and understand the system. By providing a comprehensive evaluation framework, COP-12 aims to bridge technical accuracy with user-centric design principles, making it particularly suited for assessing the explainability and trustworthiness of AI systems.

\subsection{Sociotechnical Perspectives}

Sociotechnical approaches to AI emphasize embedding systems within their broader social and organizational contexts, arguing that technological solutions alone are insufficient for addressing complex real-world challenges. Such approaches integrate human and organizational factors, fostering systems that are both technically effective and socially meaningful.

Frameworks like \textit{Critical Technical Practice (CTP)} \cite{agre1997computation} and \textit{Value-Sensitive Design (VSD)} \cite{zhu2018value} exemplify this perspective. CTP, as proposed by Agre, calls for a reflective critique of the epistemic and methodological assumptions underlying AI development. It encourages researchers to question dominant algorithmic paradigms and consider alternative designs that prioritize human values and contextual relevance. Similarly, VSD integrates stakeholder perspectives early in the design process to ensure that the systems align with the ethical and practical needs of diverse user groups.

These perspectives challenge the dominance of algorithmic formalism \cite{green2020algorithmic} the tendency to abstract AI solutions away from their real-world context. For instance, studies have highlighted how neglecting sociotechnical factors can lead to algorithmic interventions that perpetuate biases, exacerbate inequalities, or fail to address user needs effectively. By extending abstraction boundaries to include social, cultural, and organizational factors, sociotechnical approaches aim to mitigate these issues, ensuring that AI systems are not only transparent but also trustworthy and fair.

The emphasis on sociotechnical integration aligns with broader calls for \textit{localized, context-sensitive solutions} in AI development. Rather than relying on scalable, one-size-fits-all models, these approaches advocate for systems tailored to specific social and organizational settings. This localization ensures that the systems resonate with the lived experiences of their users, addressing unique challenges and opportunities within their deployment contexts.

Ultimately, sociotechnical perspectives argue for a paradigm shift in AI design—moving from algorithm-centric approaches to human-centered systems that are reflexive, inclusive, and deeply embedded within the social fabrics they aim to serve.

\section{Methods}
\subsection{Recruitment}
This study was conducted within my neighborhood and extended friend circle, leveraging snowball sampling to recruit participants.\cite{goodman1961snowball} Snowball sampling was chosen to efficiently identify individuals who fit the study's criteria, given the preliminary nature of the research and the constraints of time. Due to these constraints, flyer recruitment was not employed, as the goal was to quickly gather participants for an exploratory investigation.

Participants were specifically chosen based on their limited prior knowledge of AI systems, ensuring that the study focused on individuals with minimal exposure to or understanding of such technologies. Recruitment was verified through unscripted discussions, during which the following factors were assessed:
\begin{itemize}
    \item \textbf{Participant Identification:} Verification of eligibility for the study.
    \item \textbf{Comfort Level with AI Systems:} Ensuring participants had no intimate familiarity with AI systems.
    \item \textbf{AI System Familiarity:} Determining the type of AI systems participants were most comfortable or familiar with, if any.
\end{itemize}
Participants were asked to consent to a 30-minute discussion, during which their interactions with the system were observed and analyzed. Participants provided informed consent before participating in the study and their data has been annonymized. A total of seven participants were recruited. Two of them had some understanding of AI systems. Three of the remaining had used ChatGPT in the last six months and hence approached the AI system with that knowledge in hand. Table 1 contains the participant ID, job, and AI knowledge.
\begin{table}[H] 
\centering
\begin{tabular}{|c|l|l|}
\hline
\textbf{Participant ID} & \textbf{Job}                  & \textbf{AI Knowledge}              \\ \hline
P1                      & Data Science Student          & Medium                              \\ \hline
P2                      & Electrical Engineering Student & Low                                \\ \hline
P3                      & Data Science Student          & Medium                              \\ \hline
P4                      & Business School Major         & Low                                \\ \hline
P5                      & Retired                       & Low                                \\ \hline
P6                      & Graphic Designer              & Low                                \\ \hline
P7                      & Marketing                     & Low                                \\ \hline
\end{tabular}
\caption{Participant Codes}
\label{tab:codebook}
\end{table}

\subsection{Technical Design}
The system for this study was developed to evaluate the effects of various AI explanation types on user trust and understanding. The design integrates cutting-edge technologies to provide a flexible, interactive environment for qualitative analysis. The system consists of a React front end, an LLM-powered backend, and a Python-based explanation solver, with the following components:
\paragraph{Frontend Design}
The frontend of the system was implemented using React, enabling a user-friendly and interactive interface. This interface presents AI model predictions alongside different types of explanations. It allows users to explore explanations through dynamic explanations and textual descriptions. it also Captures user feedback and interaction data for subsequent analysis.

\paragraph{Backend and Explanation Engine}
The backend architecture leverages an LLM (Llama 3.2B)\cite{meta2024llama3_2} to enhance the system's ability to provide contextually relevant and dynamic explanations. The backend is designed to generate contextual information. The Llama 3.2B model generates natural language summaries of model predictions and contextualizes explanations based on user profiles and tasks. The model adapts explanations dynamically to fit the specific scenario or user input, tailoring outputs to enhance user comprehension.

\paragraph{Explanation Solver}
The system employs a Python-based solver to compute and present interpretable explanations. A key component is the SHAP (SHapley Additive exPlanations) package, which is used to calculate feature importance scores for each model prediction. it can then visualize how individual features contribute to the output, providing diagnostic insights for technical users and finally generate concise, non-technical explanations for users without deep technical expertise.

\subsection{AI Interaction}

Participants interacted with an AI-powered loan decision system by simulating loan applications. They provided inputs such as age, sex, requested credit amount, and income. Based on these inputs, the system would either approve or reject the loan request. After each decision, the system could provide an explanation detailing the reasoning behind its decision, while other times, no explanation was given. Participants were encouraged to submit multiple loan requests with varying inputs to explore the system's decision-making logic. Participants interacted with the system ten times with a different explanation type each time. Participants were encouraged to reflect on why the system was behaving in this manner and if the explanations generated were reasonable. 

\subsubsection{Explanation Types}
The system generates and displays four types of explanations:
\begin{itemize}
    \item \textbf{No Explanation:} No explanation is generated on the screen.
    \item \textbf{Basic Explanation:} Feature importance explanations.
    \item \textbf{Detailed Explanation:} Contextual information.
    \item \textbf{Interactive Explanation:} Dynamic user queries.
\end{itemize}

\subsection{The Interview Process}
The semi structured interviews were conducted both online over zoom and in-person. The interview had three main parts:
\begin{enumerate}
    \item \textbf{Part 1:} A casual discussion about the participants’ prior experiences with AI systems, including their perceptions and challenges, serving as a foundation for understanding their familiarity and expectations. 
    \item \textbf{Part 2:} Structured questions directly aligned with the Cop-12 metrics, targeting key aspects such as correctness, context, and confidence in explanations. While this section followed a relaxed framework to ensure all relevant metrics were addressed, the conversation was deliberately guided to encourage participants to reflect deeply on these aspects.
    \item \textbf{Part 3:} A relaxed, post-interview style, allowing participants to share their broader thoughts and fill out related feedback forms, creating an informal atmosphere to elicit candid responses.
\end{enumerate}

The entire Interview protocol can be viewed in the appendix.

the results were coded so that they aligned as much as possible with the already existing Cop-12 metrics and attempted to see if there were themes that are not captured previously. 

\subsection{Qualitative Analysis}
The qualitative analysis followed a structured approach to ensure the systematic identification of themes aligned with the Cop-12 metrics. The process began with familiarization, where interview transcripts were reviewed to understand the broad patterns and themes emerging across participant responses. Next, initial coding involved labeling specific transcript segments with descriptive codes, guided by the Cop-12 metrics. For example, one participant's statement, “The explanation didn’t reflect what the system actually did” (P1), was coded as “incorrect explanation” under the Cop-12 theme of correctness. Following this, thematic coding merged similar codes into sub-themes that remained aligned with the Cop-12 framework. For instance, codes such as “Inconsistent Outcomes” and “Incorrect Explanation” were consolidated under the sub-theme “Content Issues,” corresponding to the Cop-12 theme of content.The process of theme development involved refining and defining these sub-themes as they began to emerge more distinctly. Finally, during validation, the themes were further refined to ensure coherence and alignment with both the data and the Cop-12 metrics. This iterative process ensured that the analysis was grounded in participant responses while maintaining relevance to the established theoretical framework.\cite{braun2006thematic} Once the codes were matched with the already existing codes, further analysis could be done on the remaining themes.

\section{Findings}

The findings from the study reveal that most participant responses align well with the COP-12 framework, particularly in terms of its emphasis on content, presentation, and user-focused metrics. However, certain areas of participant feedback expose gaps in COP-12, suggesting that the framework could benefit from expanded dimensions to fully capture user needs and experiences.

\subsection{Research Question 1}
\textbf{How do different types of AI explanations shape users' initial impressions and perceived trust during their interaction with the AI system?}

Participants’ impressions of the AI system were strongly influenced by the type of explanation provided and its alignment with their expectations. Explanations that demonstrated consistency and correctness were most likely to build trust. For example, P1 noted, \textit{``When I gave 30k as income it gave me acceptance, but when I dropped it to 25k it rejected me,''} highlighting the importance of consistency in maintaining user trust. Similarly, P3 expressed skepticism when explanations lacked sufficient detail to clarify key factors: \textit{``I guess I understood that the income amount has greater weight compared to sex, but I’m not sure how much the weight is.''} These findings align with the correctness and covariate complexity sub-themes within the Content dimension of COP-12.

The type of explanation also played a role in shaping user impressions based on their level of expertise. Novice users preferred simple, contextual explanations, while participants with more experience appreciated technical details. Interactive explanations, which allowed users to query and clarify the AI’s decisions, stood out as particularly effective. P7’s question, \textit{``Can we ask it questions?''} reflects the growing expectation for interactive systems, tying directly to the controllability sub-theme under COP-12’s User-Focused Metrics. However, inconsistencies in how explanations adapted to user queries and inputs highlighted potential gaps in COP-12’s framework for evaluating dynamic, interactive systems.

\subsection{Research Question 2}
\textbf{What elements of these explanations are most influential in building or undermining users' trust during their interactions with this AI system?}

Participants identified several elements of AI explanations that influenced their trust, with transparency and contextual relevance emerging as critical factors. Transparency, particularly regarding confidence indicators, played a significant role. P4 questioned, \textit{``Is an accuracy score of 85\% even good?''} and further asked, \textit{``Is the confidence level for the final result or the explanation?''} Similarly, P7 added, \textit{``If the confidence level is high and the explanation is bad, is the system good or bad? What does the confidence level actually mean here?''} These quotes highlight the confidence sub-theme under COP-12, where users expect clear and actionable explanations of confidence values. However, the lack of clarity around confidence metrics suggests that COP-12 could benefit from more detailed guidance on presenting this information.

Contextual relevance was also crucial in fostering trust. Users valued explanations tailored to their specific decisions over generic outputs. P5 observed, \textit{``Real life is more than numbers, I would just talk to someone at the Bank''} emphasizing the need for explanations to account for real-world factors and specific decision contexts. While COP-12 captures context as a sub-theme, participants’ feedback points to gaps in the framework’s ability to evaluate how explanations adapt dynamically to user needs or queries.

A couple participants highlighted issues with accessibility and usability that undermined their trust. For example, P3 commented, \textit{``There is no way to control the length of the response the way I can do it in ChatGPT,''} reflecting the importance of compactness in ensuring explanations are concise and user-friendly. This feedback underscores the need for AI explanations to balance detail with simplicity. Accessibility challenges also extended to users’ ability to interpret technical metrics, such as accuracy. P1 remarked, \textit{``I don’t think 85\% is very high accuracy,''} while P3 added, \textit{``I think you should get the accuracy a little higher before deploying the system.''} These statements suggest that while correctness and confidence are covered in COP-12, the framework does not explicitly account for how such metrics should be presented to ensure accessibility and usability for diverse user groups.

\subsection{Tie-in with COP-12 Metrics and Limitations}

The findings demonstrate that the COP-12 framework effectively captures many dimensions critical to explainability, including correctness, consistency, compactness, and context. For example:
\begin{itemize}
    \item \textbf{Correctness and Consistency:} P1’s remark about the system’s inconsistent handling of similar income inputs highlights the importance of these sub-themes in fostering trust.
    \item \textbf{Compactness:} P3’s frustration with verbose explanations underscores the need for concise presentation formats.
    \item \textbf{Confidence:} P4 and P7’s concerns about unclear confidence metrics point to the need for transparency in this area.
\end{itemize}

However, the findings also reveal areas where COP-12 falls short:
\begin{itemize}
    \item \textbf{Fairness:} Participants questioned the ethical implications of system decisions, such as P4’s remark about the relevance of accuracy scores. COP-12 does not explicitly evaluate fairness or user perceptions of equity in AI outcomes.
    \item \textbf{Accessibility:} P3’s comments on explanation length and usability, along with P1’s and P3’s remarks on interpreting accuracy scores, highlight a gap in COP-12’s ability to evaluate whether explanations are accessible and actionable for users with varying levels of expertise.
    \item \textbf{Adaptability:} While COP-12 addresses controllability, it does not fully account for the dynamic and interactive nature of explanations that participants, like P7, found valuable.
\end{itemize}

\section{Discussion}
In the discussion section I will describe two broad themes that have come out of our analysis. 

\subsection*{Algorithmic Transparency Is Not Enough}

The findings from our study emphasize that while algorithmic transparency is a necessary step toward fostering trust in AI systems, it is far from sufficient. Participants frequently expressed skepticism about the relevance and clarity of purely technical explanations, even when these were accurate and detailed. For instance, Participant P4’s comment, \textit{``Is an accuracy score of 85\% even good?''} highlights the limitations of presenting technical metrics without sufficient contextual framing. Additionally, the lack of clarity in confidence indicators, as illustrated by P7’s confusion over what confidence scores represent, suggests that transparency efforts often fail to translate into actionable understanding for users.

Interactive explanations showed promise in bridging some of these gaps, particularly for users who were able to engage with the system through queries. However, even these explanations struggled to adapt dynamically to user needs or clarify inconsistencies in the AI’s behavior. For example, P1 noted the inconsistent handling of similar income inputs, which undermined their trust in the system. This highlights a broader issue: algorithmic transparency, as defined by metrics like correctness and compactness under COP-12, does not inherently ensure that users find the system reliable or fair.

The findings underscore the need for transparency approaches that go beyond algorithmic details. These must address user-centric dimensions such as accessibility, interpretability, and adaptability. Without these considerations, transparency efforts risk alienating users or, worse, fostering a false sense of trust in systems that fail to account for real-world complexities.

\subsection*{The Need for a Wider Social Context}

A recurring theme in participant feedback was the disconnect between the AI explanations provided and the broader social, economic, and individual contexts in which users make decisions. Participants like P5 underscored the limitations of numerical and technical outputs, remarking, \textit{``Real life is more than numbers, I would just talk to someone at the Bank.''} This sentiment reflects a critical gap in the design of current explanation systems: their inability to contextualize decisions within the lived realities of users.

Contextual explanations, which incorporate factors such as local economic conditions and individual circumstances, were highlighted as pivotal for fostering trust. However, the findings also reveal that existing frameworks like COP-12 inadequately address this need. While the framework includes \textit{context} as a dimension, it does not fully evaluate how explanations adapt dynamically to user-specific scenarios or societal factors. This limitation became evident in the feedback from participants who sought explanations that were not just technically accurate but also socially meaningful.

Moreover, participants’ concerns about fairness and accessibility point to the need for explanation systems that engage with ethical and sociotechnical dimensions of AI. P4’s questioning of the relevance of accuracy scores, for example, underscores the importance of designing systems that account for perceptions of equity and inclusivity. Similarly, P3’s frustration with verbose and inaccessible explanations highlights the need for systems that prioritize user-centric design over purely technical objectives.

By embedding explanations within a wider social context, AI systems can move beyond the limitations of algorithmic formalism to become truly human-centered. This involves not only tailoring explanations to individual users but also addressing systemic factors that influence trust, such as power dynamics, cultural norms, and economic inequalities. Only by adopting such a sociotechnical perspective can AI systems meaningfully align with the needs and expectations of diverse user groups, fostering trust in high-stakes environments.

\section{Limitations and Future Work}
There were several limitations on the study. The participant size was very small and there was no saturation in the codes. Findings from this study should not be generalized. From a technical stand point the way the interaction worked was substandard. A lot of problems might have been alleviated in case of better hardware or more investment into the LLM.

Future work will include running the experiments again with a larger sample size and a more elaborate technical set up. 

\section{Conclusion}
This study underscores the limitations of current approaches to explainability in AI systems, particularly those focused narrowly on algorithmic transparency. While technical transparency is essential, it does not inherently foster trust or ensure usability, as evidenced by participant feedback highlighting issues with confidence indicators, consistency, and accessibility. The findings reveal that effective explanations must extend beyond algorithmic details to address user-centric and sociotechnical dimensions.

Contextual explanations that incorporate societal, economic, and individual factors emerged as critical for fostering trust. By aligning explanations with the lived realities of users and addressing systemic factors such as fairness and accessibility, AI systems can move beyond their current limitations. This study advocates for a paradigm shift toward sociotechnical perspectives in AI design, emphasizing the importance of creating human-centered systems that are not only transparent but also equitable and meaningful in real-world contexts. Future work should aim to refine frameworks like COP-12 to account for dynamic, interactive, and contextually rich explanations, ensuring that AI systems effectively meet the diverse needs of their users.

\bibliographystyle{plain}
\bibliography{references}

\appendix
\section*{Appendix C: Interview Protocol}

\textbf{Introduction}

Thank you for joining this interview. We’ll be discussing your experiences with the AI system you interacted with, particularly focusing on how trust and explainability played a role in your perception of the system. There are no right or wrong answers, and your input will be invaluable.

\textbf{Broad Opening Question}

To begin, could you tell me about your overall experience with the AI system? What stood out to you, positively or negatively?

\subsection*{1. Trust in Information Quality}

\textit{Accuracy and Completeness}

How did the accuracy of the AI’s explanations affect your trust in its outputs?

Can you recall any situations where the explanation felt incomplete? How did that impact your perception?

\textit{Consistency and Continuity}

Did you notice if similar inputs yielded similar explanations? How did that affect your trust in the AI system?

Were there moments where inconsistent explanations stood out to you? What was your reaction?

\textit{Contrastivity}

Did the system address “why not?” or “what if?” scenarios effectively? Could you share examples where this worked well or fell short?

\subsection*{2. Presentation of Explanations}

\textit{Compactness and Composition}

How did the length or style of the explanations influence your trust?

Did you find concise explanations more helpful, or did detailed ones make you feel more confident?

\textit{Confidence Indicators}

How did the presence (or absence) of confidence levels or probabilities affect your trust?

What kind of information would make confidence indicators more meaningful to you?

\subsection*{3. User-Centric Design}

\textit{Relevance and Context}

Were the explanations relevant to your needs or the task you were trying to accomplish? Can you give an example?

What do you think the AI could do to make its explanations more useful for you?

\textit{Coherence and Controllability}

How well did the explanations align with your existing knowledge or expectations? Could you describe moments where this worked well or didn’t?

Did you feel in control of the AI’s explanation process? What features would enhance that sense of control?

\textbf{Conclusion}

Wrap up:

Is there anything else about your experience with the AI system that significantly affected your trust or understanding that we haven’t discussed?

\end{document}